\documentstyle[prl,aps,epsf]{revtex}
\begin{document}
\draft
\twocolumn[\hsize\textwidth\columnwidth\hsize\csname @twocolumnfalse\endcsname

\title{Superconducting Fluxon Pumps and Lenses} 
\author{J.F.~Wambaugh$^1$, C.~Reichhardt$^2$, C.J.~Olson$^2$, 
F.~Marchesoni$^{1,3}$, 
and Franco Nori$^{1}*$}
\address{$1. \ $ Department of Physics, 
The University of Michigan,
Ann Arbor, Michigan 48109-1120 \\
$2. \ $ Physics Department, University of California, Davis, CA 95616 \\
$3. \ $ Istituto Nazionale di Fisica della Materia, Universit\'a di Camerino, 
Camerino, I-62032, Italy 
}

\date{\today}
\maketitle
\begin{abstract}
We study stochastic transport of fluxons in superconductors by 
alternating current (AC) rectification.  Our simulated system  
provides a fluxon pump, ``lens", or fluxon ``rectifier" 
because the applied electrical AC is transformed into a 
net DC motion of fluxons.  
Thermal fluctuations and the asymmetry of the ratchet channel 
walls induce this ``diode" effect, which can have important 
applications in devices, like SQUID magnetometers, and for 
{\it fluxon optics}, including convex and concave 
{\it fluxon lenses}.  
Certain features are unique to this novel two-dimensional (2D) 
geometric pump, and different from the previously studied 
1D ratchets.
\end{abstract}
\vspace{-8pt}
\vspace*{-0.3in}
\vskip2pc]
\narrowtext

{\it Introduction.---\/}
A number of authors (see, e.g., \cite{intro1,intro2,fabio,doering})
have recently addressed the longstanding problem of how to extract 
useful work from a fluctuating environment.  
While heat may not be transformed back to mechanical 
work at thermal equilibrium 
(i.e., without temperature gradients \cite{feynman}), 
the same restriction does not apply to the case of 
nonequilibrium thermal fluctuations:  An asymmetric device 
(like Feynman's ratchet \cite{feynman}) can indeed rectify 
symmetric quasi-equilibrium fluctuations 
\cite{intro1,intro2,fabio,doering}.  
The implications of such a mechanism in transport theory are 
far-reaching:  macroscopic currents may arise even in the absence 
of external forces or gradients.

Recently, a few groups \cite{zapata,nd} have made 
interesting proposals for quite distinct ways of 
using ratchets in superconductors.  
Ref. \cite{zapata} uses Josephson junctions, in SQUIDs and 
arrays, while Ref.\cite{nd} uses a {\it standard 1D\/} 
potential-energy ratchet (e.g., \cite{intro1,fabio,doering})
to drive fluxons out of superconducting samples 
using simulations like in \cite{simulations,paps}.
Here, we use the new concept of {\it 2D asymmetric channel walls\/} 
to provide an experimentally-realizable way to easily move fluxons 
through channels in AC-driven devices.  These novel 
{\it geometric ratchets\/} would allow, for example, 
the construction of {\it fluxon optics\/} devices,  
including concave/convex fluxon {\it lenses} 
that concentrate/disperse fluxons in nanodevices.
Further, we also consider the effects of 
{\it collective\/} interactions on ratchets, 
instead of the usual {\it one--particle\/} 
stochastic transport.

{\it Simulation.---\/}
Computer simulations were performed using a new version of 
the molecular dynamics (MD) code used for systems with random 
\cite{simulations} and correlated \cite{paps} pinning. 
We model a transverse 2D 
slice (in the $x$--$y$ plane) of an infinite zero-field-cooled 
superconducting slab containing current-driven 3D rigid
vortices that are parallel to the sample edge.  The samples 
have very strong, effectively infinite, pinning except the 
``zero pinning" {\it central saw-tooth-shaped channel\/} 
(see inset of Fig.~1).  In the latter, fluxons moved 
subject to:  fluxon-fluxon repulsions ${\bf f}_{vv}$, 
an externally-applied AC Lorentz driving force 
${\bf f}_L$, forces due to thermal fluctuations ${\bf f}_T$, 
and interactions with the channel boundaries,   
treated as infinite potential barriers. 
The thermal force ${\bf f}_T$ was implemented 
using a Box-Muller zero-mean, Gaussian 
distributed random number generator.
We measure all forces in units of 
$f_{0}=\Phi_{0}^{2}/8\pi^{2}\lambda^{3}$,
magnetic fields in units of $\Phi_{0}/\lambda^{2}$, 
and lengths in units of the penetration depth $\lambda$.
Here, $\Phi_0 = hc/2e$ is the flux quantum.
Our dimensionless temperature is
$T = k_B T_{\rm actual} / \lambda f_0$.  
Thus, away from the channel edge, 
the total force on a fluxon is 
$ {\bf f} = 
{\bf f}_{vv} + {\bf f}_{L} + {\bf f}_{T} = 
\eta{\bf v}$,
where the total force on vortex $i$ due to other vortices  
is given by
${\bf f}_{vv}^{(i)} $
$= \sum_{j=1}^{N_{v}}f_{0} \ K_{1}(|{\bf r}_{i} - {\bf r}_{j}|/
\lambda) \, {\bf {\hat r}}_{ij} .$
Here, 
$K_1$ is a modified Bessel function, 
${\bf r}_{i}$ (${\bf v}_{i}$) is the location (velocity) of the $i$th vortex,
$N_{v}$ is the number of vortices, 
${\bf {\hat r}}_{ij} =({\bf r}_{i} - {\bf r}_{j})/|{\bf r}_{i} - {\bf r}_{j}|$, 
and we take $\Delta t = 0.01$ and $\eta = 1$.  
Simulations for the same parameters were repeated with 
different random number seeds, to provide for disorder averaging.

Unless otherwise specified, each figure refers to simulations 
conducted in the following way:  Initially fluxons were randomly 
placed inside the channel and subjected to an alternating current 
along $y$---producing a square-wave Lorentz driving force along 
$x$ with $F \equiv f_L / f_0 = 15 \,$ and period  $P = 2 \tau$ 
(here, $\tau$ = 100 MD steps).  This means that in the absence 
of thermal noise ($T=0$) and channel walls, a single fluxon 
would alternate traveling $15 \lambda$ in one direction 
(e.g., $+x$), and then $15 \lambda$ in the other ($-x$).  
The period $9 \lambda$ of the horizontal ratchet teeth was such 
that the driving force $F=15$ should be sufficient to allow for 
a rectification, or diode effect, because it moves the 
fluxon back and forth through the bottlenecks indicated in the 
inset of Fig.~1.  Ten runs, of $250 000$ MD steps each, 
were used to find the average current and standard deviation for 
each plotted point.  When a standard deviation is not shown, it was 
significantly smaller than the plotted point.  All samples had a 2D
geometric ratchet made of asymmetric walls.  The simulated sample 
was an $18 \lambda$ by $18 \lambda$ square, with periodic boundary 
conditions.  The channel is $7 \lambda$ wide, with 
saw-teeth of period $9 \lambda$ (four teeth per unit cell) and 
slope $1/3$ (except in Fig.~5, and in other runs to be discussed 
elsewhere, where geometry was varied).  This 
leads to a bottleneck that is one $\lambda$ wide.  This 
construction is a novel 2D ratchet that uses geometry, 
which governs the system potential energy, to cause rectification.

{\it Temperature dependence of the fluxon 
rectification.---\/}Figure 1 shows the rectified fluxon velocity 
$ \langle v \rangle = \sum_i^{N_v} v_i / N_v \;$ versus 
temperature.  The dimensionless temperature $T$ was 
varied over two orders of magnitude, from $1$ to $100$.   
These simulations clearly indicate an optimal or ``resonant" 
temperature regime in which the DC fluxon velocity is maximized 
by the fluxon pump or diode.  This optimal temperature regime 
can be explained as a trade-off between allowing the fluxons 
to fully explore the ratchet geometry (i.e., $T$ must not be
too low) and washing out the driving force (i.e., $T$ must 
not be too large).  At low temperatures, the alternating driving 
force will cause a fluxon to migrate to the center of the channel 
and no longer be impeded or assisted by interactions with 
the geometry; while at high temperatures the driving force 
becomes irrelevant and thus the fluxon is no longer pushed 
through bottlenecks regularly.  The $\langle v \rangle (T)$ 
curve for many randomly placed fluxons is similar to the single 
fluxon case, but the magnitude of $\langle v \rangle (T)$ 
decreases when the fluxon density $B$ increases.
This is because the repulsive force produced by a large number 
of fluxons act to restrict each other's motion.  Note that 
at the optimal temperature, the fluxons travel nearly 
$15 \, \lambda$ every $100$ MD steps.  The general shape 
of the curve was established with two fine scans with one 
and fifty fluxons.  Conformity of other field strengths 
to this curve was judged by performing a coarser scan, 
then connecting the points with a spline.
The small dip in the $\langle v \rangle$ for low $T$ 
will be discussed separately (see Fig.~3 below).

{\it Driving period dependence of the flux pump.---\/}  
Fig. 2(a) shows $ \langle v \rangle (P)$ for six different 
combinations of driving force amplitudes $F$ and temperatures 
$T$ for a single fluxon.  Interestingly, as the period $P$ 
of the driving force was varied from low to high (from 
$P = 20$ MD steps to $2000$ MD steps) {\it no optimal peak\/}  
in $ \langle v \rangle$ was discovered.  This is in contrast 
with previous work on ratchets, which provide a peak in 
$ \langle v \rangle$ versus either $P$ or frequency.  Instead, 
we find that 
while a high frequency driving force 
yields a very small ratchet velocity, the 
velocity quickly converges to a nearly stable value with 
increasing $P$.  This result corresponds to the idea that 
a fluxon must be forced through the bottleneck of the 
geometric 2D ratchet in order for it to be a rectifier.  
At high frequencies, a fluxon does not travel far enough to 
interact with the bottleneck.  Once the frequency is low 
enough to force it through a bottleneck, however, the net
velocity changes little by forcing it through further 
bottlenecks.  

All of the simulations in Fig.~2(a) were conducted at 
relatively low temperatures to allow the effects of the 
driving force to dominate.  In addition to trying 
$F = f_L / f_0 =  15$, the value typically used in 
this paper, two combinations of driving force and 
temperature magnitudes were tried that both also had an 
$F/T$ ratio of $3/1$.  While varying $P$ at the 
low-$T$ regime had little effect upon the period 
dependence curve, varying $F$ very clearly did have 
an effect:  The higher $F$, the greater $\langle v \rangle$.

Fig.~2(b) shows the dependence of $\langle v \rangle$
on $P$, for $T = F / 3 = 5$ with many fluxons instead of one.  
Initially, fifty randomly placed fluxons were simulated.
As with the single fluxon case, a high frequency driving force 
resulted in a low $\langle v \rangle$ because the fluxons were 
not being forced through the ratchet bottleneck.  Increasing 
$P$ produced a rapidly plateauing $\langle v \rangle$.  
Once the shape of the curve was determined, similar sets of 
observations were made of different fluxon densities $B$ at six 
different periods.  These simulations of many fluxons 
demonstrate that, as with one fluxon, there was 
{\it no\/} optimal peak.

{\it Fluxon pump effect versus field.---\/}By varying the 
density of fluxons $B$, we found a maximum in $\langle v \rangle$
at very low temperatures (see Fig.~3). This is due to an enhanced 
interaction with the ratchet geometry: at extremely low $T$ 
the driven fluxons eventually work their way to the middle 
of the channel and then no longer interact with 
the ratchet.  A small increase in $B$ can drive fluxons 
from the middle of the channel, causing these fluxons to 
increase their interaction with the sawteeth, without 
significantly increasing the resistance due to fluxon-fluxon 
interactions.  Aside from this low-$T$ resonance, 
increasing $B$ slowly decreases $\langle v \rangle$, 
because, as the fluxons repel one another, at high densities 
they {\it block\/} the bottleneck.  
At high temperatures (e.g., $T=50$) 
the ratchet velocity decreases slowly with increased fluxons.  
In this case, the slight maximum in $\langle v \rangle$ found 
at low $T$ is not present because the resistance due 
to fluxon-fluxon interactions is always larger than the gains 
from increased exploration of the geometry of the ratchet.

{\it Flux pump response to the driving force.---\/}Increasing 
the amplitude $F$ of the square-wave driving force increases 
the average velocity $ \langle v \rangle$ 
of a fluxon in the ratchet, as shown in Fig.~4.  This 
monotonic increase in $ \langle v \rangle$ versus 
$F$ for 2D geometric ratchets is different from the peak 
observed in 1D potential ratchets (e.g., see Fig.~2 of \cite{nd}).  
This makes it fairly easy for any particular 2D geometric ratchet 
to be continuously tuned to desired values of $ \langle v \rangle$.  
This would be an advantage of 2D geometrical over 1D potential 
ratchets for potential practical designs of devices.  A reason 
for their different response is because the resistance to 
motion due to successive bottlenecks does not add like the 
resistance due to successive potential barriers.     
If a fluxon happens to remain for some time in 
the center of a bottleneck, unlikely to happen at high enough 
temperatures, then it will not be rectified; while the 
potential barrier rectifies at every period.  Increased 
fluxon density again increases resistance, reducing 
$ \langle v \rangle$.

{\it Geometry dependence.---\/}To determine the 
generality of the simulation results presented here with respect 
to varied ratchet geometries, in Fig.~5 we compare 
$\langle v \rangle (T)$ for fifty fluxons in two alternate 
geometries with the $\langle v \rangle (T)$ shown in Fig.~1 
(open squares in both figures).  Increasing the slope, and 
correspondingly decreasing the period, of the saw-teeth 
clearly shifts the optimal temperature to lower values.  
Despite this shift, however, the qualitative behavior of the 
first geometry studied carried over to other ratchet slopes 
and periods.  Additional samples and parameters, including 
wider bottlenecks, were also studied (but not discussed here 
due to space limitations), giving consistent results.

{\it Discussion.---\/}
Our 2D geometric ratchet has some properties which are similar 
to previously studied 1D potential ratchets.  
Like the simpler 1D cases, the 2D ratchet showed a 
``resonance region" with a maximum in $\langle v \rangle$, and 
(outside of a low-$T$ anomaly) a decreased $\langle v \rangle$ 
with increasing $B$.   Unlike the 1D cases, however, the ratchet 
velocity plateaus with decreased driving frequency, 
instead of displaying a peak.  
Also, 2D geometric ratchets are easier to implement 
experimentally since these do not require carving 
very many inclined asymmetric grooves, as with a 
1D potential ratchet \cite{nd}.  
Moreover, as pointed out above (fig.~4), 
$\langle v \rangle$ can be continuously tuned by 
varying $F$, for the 2D geometric ratchets.

{\it Concluding remarks.---\/}Superconducting devices using 
stochastic transport were studied because they provide a 
convenient way to move fluxons through channels.  At non-zero 
temperatures, the asymmetric geometry of the ratchet saw-teeth 
automatically converts applied AC inputs into a net DC motion 
of fluxons.  Thus, the simulated device serves as an excellent 
fluxon rectifier. 
Our device could be easily made experimentally 
either by electron beam lithography, irradiation, 
evaporating layers, or by chemically etching a 
channel.  The central channel would have very weak 
pinning, while the rest very strong pinning.
By coupling two ratchets that rectify in opposite 
directions (see inset of Fig.~2(b) and \cite{magnifiedfigs}), 
fluxon {\it lenses\/} that could 
either (c) {\it disperse\/} or (d) {\it concentrate\/} 
fluxons in chosen regions of a sample can be created.  
More complex {\it fluxon optics\/} microdevices could be 
built similarly.  Corners (e) can also be constructed. 
Such remarkable devices, and modifications of them, 
would allow the transport of fluxons along complicated 
nanofabricated channels:  a microscopic network of fluxon 
channels in superconducting devices.  
This could be very useful to get rid of unwanted, trapped 
flux in SQUID magnetometers, and also to move fluxons 
along channels in devices \cite{squids}.  These 
promising concepts are largely unexplored, and 
constitute an open and potentially useful area.

CJO (JFW) acknowledges support from the GSRP of the 
microgravity division of NASA (NSF-REU).  We thank 
R.~Riolo and the UM-CSCS for providing computing resources.  
We acknowledge conversations with C.~Doering, 
S.~Field, A.~Barabasi, and E.H. Brandt.

\begin{figure}
\caption{
Rectified average fluxon velocity  
$\langle v \rangle$, 
which can be measured as a voltage, 
versus temperature $T$ for the ratchet geometry in the inset.  
The magnetic field $B$ is directed out of the figure.  
${\bf J} = J {\bf \hat{y} }$ is a vertically applied 
alternating current square-wave that drives the fluxons 
back and forth horizontally along the channel.
The number of fluxons is: 
1 ($\bigcirc$), 25 ($\Diamond$), 50 ($\Box$),  
75 ($\triangle$),  100 ($\times$),  150 ($+$), 
250 ($\ast$).  
}
\label{fig:v-temp}
\end{figure}

\begin{figure}
\caption{
Average fluxon velocity $ \langle v \rangle$, 
versus driving force period $P$ for 
(a) one fluxon and (b) many fluxons. For very low 
periods little rectification occurs, and for higher 
periods, $ \langle v \rangle$ slowly increases.
The four $F=15$ curves in the middle of (a) 
show similar behavior and correspond to 
(top to bottom at $\tau=10$):
$T=1$ ($\bullet$),  
$T=2$ ($\Diamond$),
$T=3$ ($\times$),
$T=2$ ($+$).
In (b), similar $ \langle v \rangle$ occur for
$50$ ($\Diamond$), 
$100$ ($\Box$), 
$150$ ($\triangle$), and
$250$ ($\times$) fluxons.
Inset: 
Schematic diagram for concave (c) and convex (d) 
{\it fluxon lenses\/} that disperse/focus fluxons 
from/at their centers. (e) shows a corner unit.
}
\label{fig:v-force}
\end{figure}

\begin{figure}
\caption{
$ \langle v \rangle $ versus the number of fluxons, 
for several values of the temperature $T$.  
}
\label{fig:v-fluxons}
\end{figure}

\begin{figure}
\caption{
$ \langle v \rangle $ versus magnitude $F$ of the driving force.
The number of fluxons for the low-$T$ ($T=5$) cases are: 
1 ($\bigcirc$), 
50 ($\Diamond$), and
100 ($\Box$).
$50$ fluxons were used for the high-$T$ ($T=50$) curve ($\triangle$). 
}
\label{fig:v-force2}
\end{figure}

\begin{figure}
\caption{
$ \langle v \rangle $ versus $T$ for different ratchet geometries, 
characterized by $(S, P_{st})$, 
where $S$ is the slope with respect to $x$, and 
$P_{st}$ the spatial period of the sawtheeth. 
The curves shown correpond to 
$(1/3, 9)$ ($\Box$), 
$(2/3, 9/2)$ ($\triangle$), and
$(1, 3)$ ($\bigcirc$). 
}
\label{fig:v-geo}
\end{figure}

\end{document}